\documentclass[ reprint, achemso, apl, two column, showkeys, groupedaddress]{revtex4-1}
\usepackage{graphicx,epsfig}
\usepackage{amssymb}
\usepackage{amsmath}
\usepackage{bm}
\usepackage{textcomp}
\usepackage{soul,xcolor}
\setcitestyle{super}

\usepackage{color}

\begin{document}
\setcounter{page}{1}
\setstcolor{red}

\title[PL]{Spatial Mapping of Local Density Variations in Two-dimensional Electron Systems Using Scanning Photoluminescence}
\author{Yoon Jang \surname{Chung}}
\email{edwinyc@princeton.edu}
\author{Kirk W. \surname{Baldwin}}
\author{Kenneth W. \surname{West}}
\author{Nicholas \surname{Haug}}
\author{Johannes \surname{van de Wetering}}
\author{Mansour \surname{Shayegan}}
\author{Loren N. \surname{Pfeiffer}}

\affiliation{Department of Electrical Engineering, Princeton University, Princeton, NJ 08544, USA  }

\begin{abstract}

We have developed a scanning photoluminescence technique that can directly map out the local two-dimensional electron density with a relative accuracy of $\sim2.2\times10^8$ cm$^{-2}$. The validity of this approach is confirmed by the observation of the expected density gradient in a high-quality GaAs quantum well sample that was not rotated during the molecular beam epitaxy of its spacer layer. In addition to this global variation in electron density, we observe local density fluctuations across the sample. These random density fluctuations are also seen in samples that were continuously rotated during growth, and we attribute them to residual space charges at the substrate-epitaxy interface. This is corroborated by the fact that the average magnitude of density fluctuations is increased to $\sim9\times10^{9}$ cm$^{-2}$ from $\sim1.2\times10^9$ cm$^{-2}$ when the buffer layer between the substrate and the quantum well is decreased by a factor of seven. Our data provide direct evidence for local density inhomogeneities even in very high-quality two-dimensional carrier systems.  

\begin{keywords} {Molecular beam epitaxy, Photoluminescence, Two-dimensional electron system (2DES), Quantum well, Electron density mapping}
\end{keywords}

\end{abstract}
\maketitle


Low-disorder two-dimensional electron systems (2DESs) are fundamental components of modern solid state physics research. Long-scale ballistic transport is essential for the operation of quantum devices, and many-body-driven phenomena typically emerge in samples having extremely high quality. Pioneering works to achieve clean 2DESs were performed in GaAs/Al$_{x}$Ga$_{1-x}$As heterostructures \cite{Stormer,Gossard,Shayegan.APL,Pfeiffer.APL,DW1}, and they have led to extraordinary achievements such as the discovery of the fractional quantum Hall effect \cite{Tsui.PRL.1982}, the observation of ballistic transport \cite{QPC,QPC2}, and the realization of exotic many-body states such as a Wigner crystal \cite{Wigner1,Wigner2,Wigner3} and nematic/stripe phases \cite{stripebubble,Fradkin.Review}. The motivation to obtain better quality 2DESs has also extended to other single-crystal material systems, and several breakthroughs have made electron-electron interactions observable in 2DESs prepared in Si \cite{Si1,Si2}, AlAs \cite{Etienne.APL,Chung.PRM}, GaN \cite{GaN}, CdTe \cite{CdTe}, ZnO \cite{ZnO}, Ge \cite{Ge}, and InAs \cite{InAs}. Recent advancements in the quality of 2D materials such as graphene has yielded phenomenal results as well, demonstrated by the observation of unconventional even-denominator fractional quantum Hall states \cite{Zibrov.NatPhys} and correlated insulating \cite{Cao.Nature1} and superconducting \cite{Cao.Nature2} phases.

Inspired by the fruitful outcome of such improvements to 2DES quality, endeavors to fabricate even higher quality samples continue to this day. For example, the primary source of scattering in state-of-the-art GaAs 2DESs is often attributed to background impurities \cite{Shayegan.APL,Pfeiffer.APL}, and to this end there have been several reports on methods to purify the source material \cite{ManfraGa,Gaproblem,Alproblem}. Another possible scattering source is the presence of density inhomogeneities within the 2DES, which have been speculated to cause unwanted effects during measurement \cite{LMR1,DensityInhomogeneity,Map,LMR2,quantumlifetime}. However, in general electrical measurements are too global to detect these inhomogeneities. Previously, we have demonstrated the capability to measure the local 2DES density of a GaAs sample at a fixed location on the wafer using a photoluminescence (PL) technique \cite{Kamburov.APL}. Here, we show that we have further developed this method and can now spatially map out the local 2DES density variations over macroscopic regions on the wafer with better than 100 $\mu$m precision. In contrast to earlier studies \cite{LMR1,DensityInhomogeneity,Map,LMR2,quantumlifetime} where the average density variations were inferred from transport measurements over the entire device, our approach provides a \textit{direct} and \textit{local} probe of the 2DES density. We find that density fluctuations on the order of $\sim10^9$ electrons cm$^{-2}$ exist even in high-mobility samples designed to have uniform 2DES density, which quantitatively supports some of these previous reports \cite{LMR1,quantumlifetime}.


\begin{figure*} [t] 
\centering
    \includegraphics[width=1.0\textwidth]{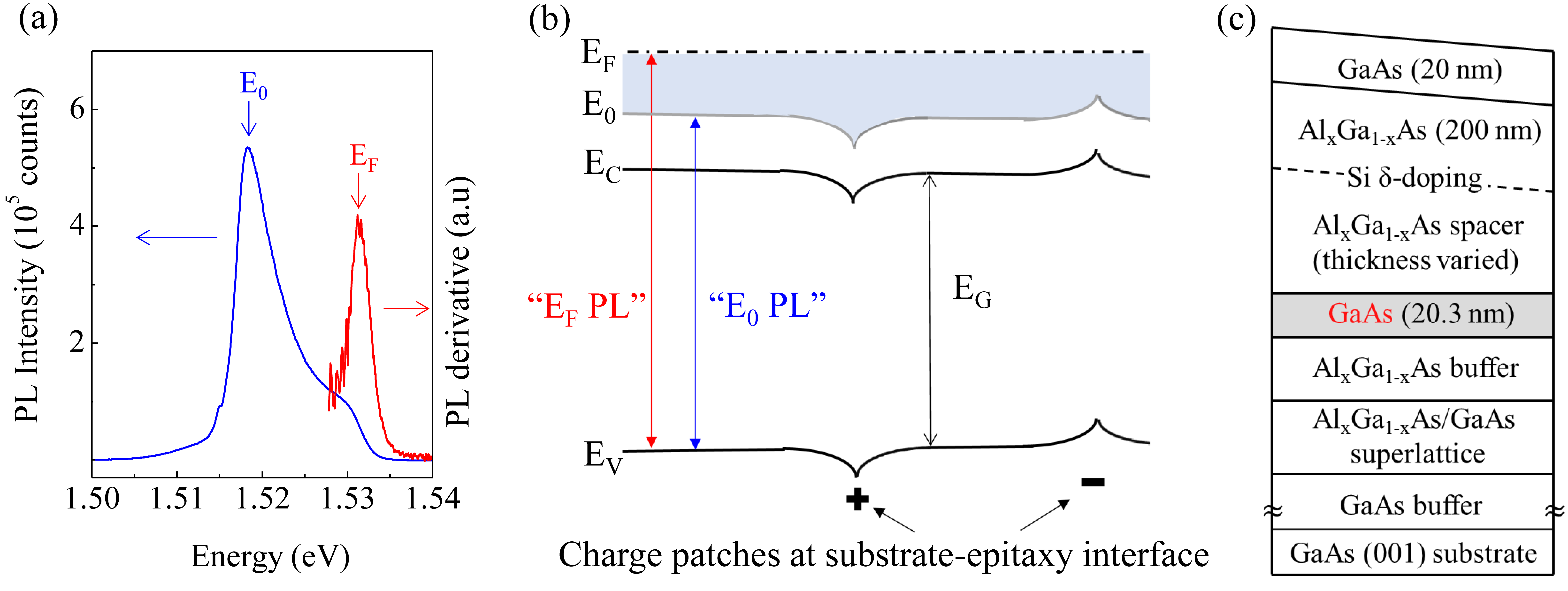}

 \caption{\label{fig1} (a) Representative PL spectrum of our GaAs 2DESs, taken at $T\simeq7$ K. The blue trace shows the raw data while the red curve shows a digitally-determined derivative of the PL spectrum. (b) Schematic energy diagram of the PL measurements. All energies are defined relative to the top of the valence band $E_V$, and the local 2DES density is derived from the difference between the Fermi energy, $E_F$, and the ground state energy of the quantum well, $E_0$. Any anomalous charge in the sample will bend the energy bands accordingly and cause local density variations. (c) Layer structure of a sample which was not rotated during the growth of the spacer layer, leading to a 2DES density gradient.}
\end{figure*}

\begin{figure*}[t]
 
 \centering
    \includegraphics[width=1.0\textwidth]{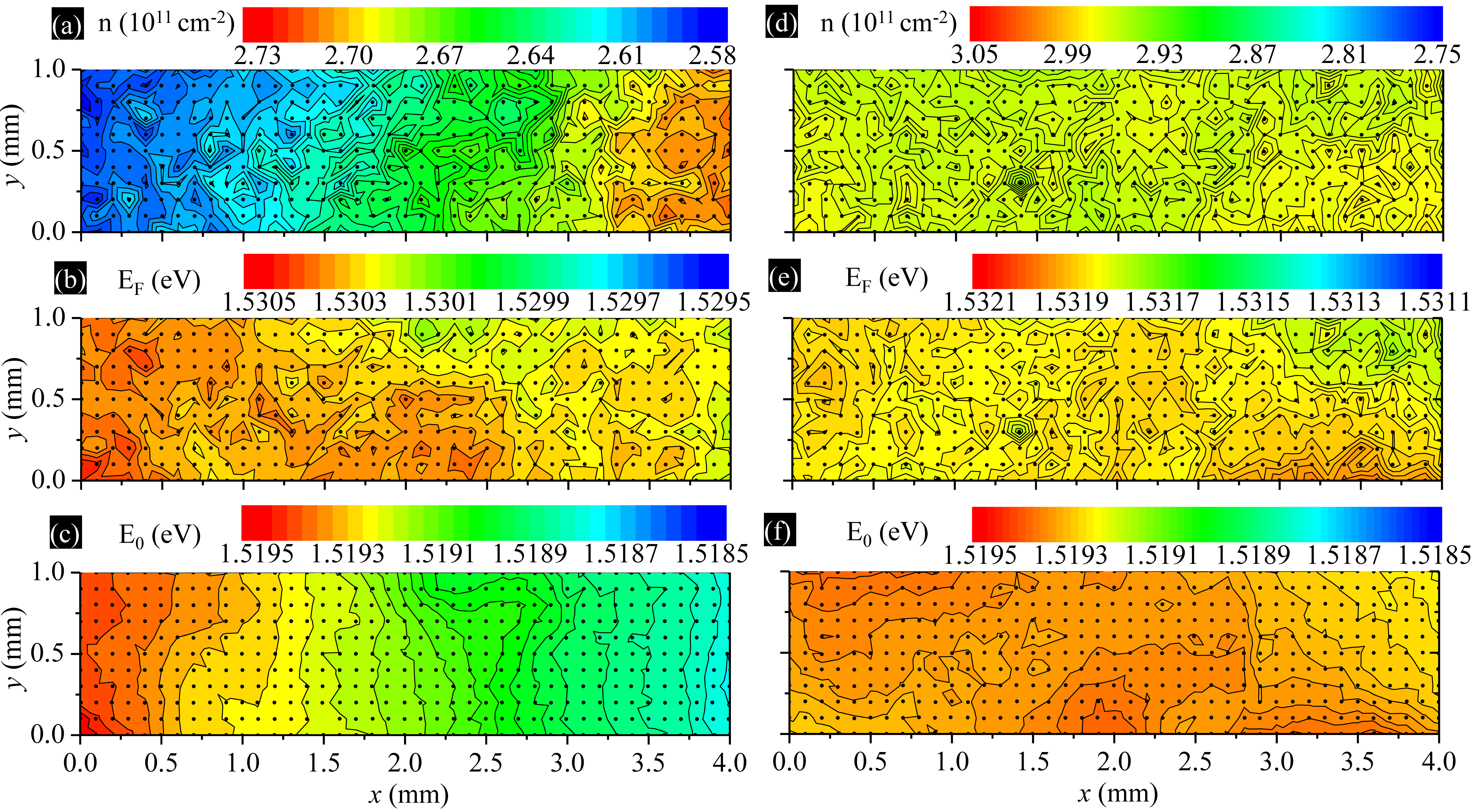} 
  \caption{\label{fig2} PL measurement contour maps of 1 mm$\times$4 mm samples. Each black dot depicts a point where the PL spectrum was measured and analyzed. The spacing between points is 100 $\mu$m along both axes. For (a)-(c) the sample was not rotated during the growth of the spacer layer, while (d)-(f) show data for a sample that was rotated at a rate of 10 RPM during the entire growth period. The 2DES density deduced from the difference between the measured $E_F$ and $E_0$ values is shown in (a) and (d). $E_F$ and $E_0$ are plotted in (b), (c) for the non-rotated sample, and in (e), (f) for the rotated sample. As illustrated in (d), density inhomogeneities are quite noticeable even in high-quality samples designed to have a uniform 2DES density. All PL spectra were taken at $T\simeq7$ K.}
\end{figure*} 

\begin{figure}[t]

\centering
    \includegraphics[width=0.45\textwidth]{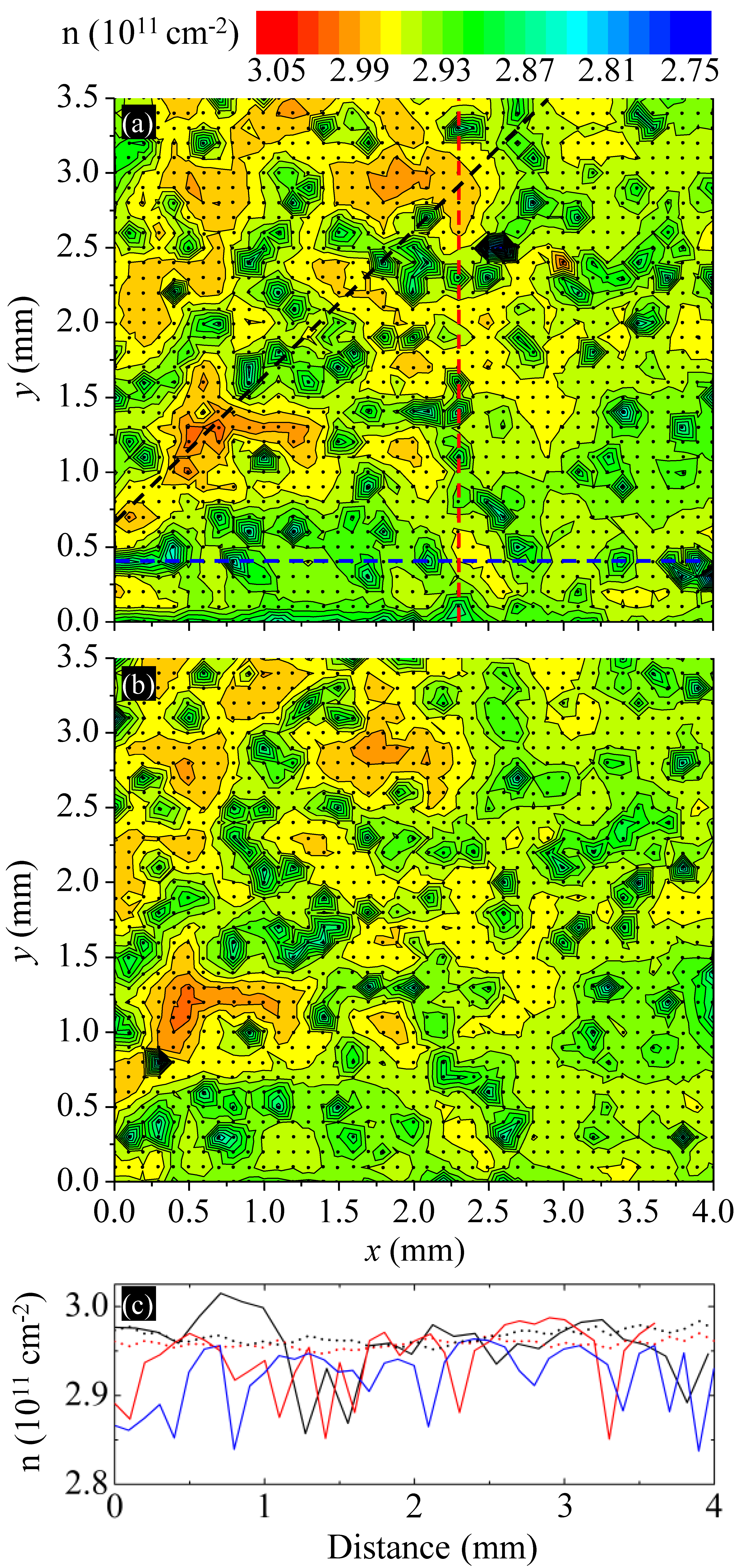} 
  \caption{\label{fig3} (a) Contour map of local 2DES density for the 200-nm-thick buffer sample. (b) Density map of the same region shown in (a) but after a full thermal cycle. (c) Line profiles of the density contour maps shown in Figs. 3(a) and Fig. 2(d): The colored solid lines correspond to the dashed lines shown in Fig. 3(a) while the black (red) dotted lines correspond to line profiles along the $x$-axis at $y=100$ (900) $\mu$m in Fig. 2(d). All PL spectra were taken at $T\simeq7$ K.}
\end{figure} 

As described in detail in our previous work \cite{Kamburov.APL}, our experiment is set up to extract the local 2DES density of GaAs quantum wells (QWs) placed in a continuous flow cryostat using PL. A 730-nm laser with an intensity of 1.2 $\mu$W focused on a spot size with a diameter of 40 $\mu$m was used to generate the PL signal. In addition, to achieve spatial mapping, the samples were mounted on a contact-cooled, piezoelectric stage whose in-plane movement can be controlled with sub-micrometer precision (Attocube ANC350). In our studies, the minimum spacing between measurements was 100 $\mu$m. All measurements were performed at the base temperature of our apparatus, which was roughly 7 K.  Our GaAs QW samples are grown on (001) oriented GaAs substrates by molecular beam epitaxy (MBE), with Si $\delta$-doping only on the surface side of the structure to have an electron density of $n=2.9\times10^{11}$ cm$^{-2}$ and a mobility of $\mu=6.3\times10^6$ cm$^{2}$ V$^{-1}$s$^{-1}$. A relatively narrow well width of $w=20.3$ nm was chosen to prevent any possible complications in the Fermi-edge from second sub-band occupation. The thickness of the buffer layer between the substrate and QW is fixed to 1.4 $\mu$m, except for the case where an intentionally thin buffer of 200 nm was implemented to amplify effects from the substrate. Figure 1(a) shows a representative PL spectrum of such a GaAs 2DES measured in our system. Gaussian fits to the raw data and the derivative are used to determine the energy positions of the peaks, which correspond to the ground state energy ($E_0$) and Fermi energy ($E_F$), respectively \cite{Kamburov.APL}.

Before discussing our data, we describe the terminology used in our measurements. As schematically depicted in Fig. 1(b), all the energy levels, including $E_0$ and $E_{F}$ are defined with respect to the valence band $E_V$ of GaAs. The 2DES density is deduced by subtracting $E_0$ from $E_{F}$ and mutliplying by the density of states [$n=(E_{F}-E_0)\times(m^{*}/\pi\hbar^2)$]. Because the focused spot size of our laser is only 40 $\mu$m in diameter, we can map out the 2DES density on a local scale as the spot is scanned across the sample. For instance, if there are any positively (negatively) charged impurities in the vicinity of a specific area excited for a PL measurement, the spectra would show an increase (decrease) in $E_{F}$ and hence the 2DES density, as shown in Fig. 1(b). The relative accuracy of our 2DES measurements is $\sim2.2\times10^8$ cm$^{-2}$, which is determined by sampling the local 2DES at a fixed point on the wafer 10 consecutive times and taking the standard deviation. This is comparable to the difference between the local 2DES density values measured at the same position, with a measurement at another point on the wafer in between, which is $\sim2.9\times10^8$ cm$^{-2}$.

In order to verify the validity of our approach, we first grew a GaAs sample that was designed to have a 2DES density gradient of roughly 1\%/mm. This was achieved by utilizing the fact that our source ovens are pointed at a 60{\textdegree} angle to the sample growth plane, and stopping the rotation of the wafer only during the growth of the AlGaAs spacer layer. The spacer thickness is therefore varied across the wafer, leading to a density gradient. All other parts of the sample, including the GaAs QW, were grown at a fixed rotation rate of 10 RPM. A schematic diagram that depicts the layer structure of this sample is shown in Fig. 1(c). Figure 2(a) shows a 2DES density contour plot of a 4 mm$\times$1 mm piece of the non-rotated sample. There is a clear density gradient observed across the long axis of the sample, which corresponds to $\sim4.5$\% over 4 mm. This is quite close to the 1\%/mm variation we expected from our growth conditions, which demonstrates that our approach to spatially map out the density of a 2DES using PL is indeed viable.

A closer inspection of Fig. 2(a) shows that, in addition to the expected global variation of density over the sample, there are also smaller, more local fluctuations in the 2DES density. To analyze this in finer detail we individually examine the spatial variations of the two terms used to deduce the density, $E_F$ and $E_0$, as shown in Figs. 2 (b) and (c), respectively. The fluctuations of $E_F$ shown in Fig. 2(b) seem quite random in both position and intensity while the variation of $E_0$ shown in Fig. 2(c) is smooth and continuous. The gradual decrease observed in $E_0$ as the 2DES density increases along the $x$-axis coincides well with the anticipated effects from band-gap renormalization \cite{BGR}. Comparing Figs. 2(b) and 2(c), it is clear that most, if not all, of the local fluctuations observed in the density [Fig. 2(a)] come from $E_F$.

This is further confirmed in the PL measurements of a sample that was rotated during the entire growth period, as illustrated in Figs. 2(d)-(f). Figure 2(d) shows the density map of the rotated sample. The global density gradient is absent in this sample, but there are still obvious local density inhomogeneities. The data shown here provide a direct measurement of density inhomogeneities in a high-quality GaAs 2DES. In the past, such inhomogeneities have only been inferred indirectly from transport measurements over the bulk of a device \cite{LMR1,DensityInhomogeneity,Map,LMR2,quantumlifetime}. Similar to the non-rotated case, in Fig. 2(e) $E_F$ for the rotated sample shows noticeable local structure with random fluctuations sprinkled throughout the sample. This time, the $E_0$ values plotted in Fig. 2(f) show very little variation across the measured area, consistent with the sample being designed to have a uniform 2DES density. We speculate that the step-like features in Fig. 2(f) come from layer fluctuations during the growth. From Figs. 2(e) and (f), it is again evident that the local density fluctuations mainly stem from changes in $E_F$. As mentioned earlier, because $E_F$ in our experiments is defined from the valence band, the data then imply that there are local charge patches scattered throughout the sample that generate electric fields and result in a non-uniform 2DES density.


We attribute these charge patches mainly to residual impurities at the substrate-epitaxy interface. It is well known that it is difficult to completely remove certain elements such as C, O, and Si from the surface of commercial GaAs substrates even after extended periods of in-situ thermal desorption prior to MBE growth \cite{Saito.J.Crys.Growth,Lee.J.Crys.Growth}. Alternatively, because the PL measurements were not performed in-situ, there could be unwanted adatoms from the atmosphere that act as either donors or acceptors on the GaAs surface. Residual impurities incorporated in the structure during the MBE growth, as well as crystalline defects such as oval defects or dislocations in the GaAs substrate, may also impact the local 2DES density.


In order to gain more insight on the origin of the density fluctuations, we grew a GaAs sample that has identical spacer thickness and QW width as the sample of Fig. 2(d) but with a factor of 7 thinner buffer layer of 200 nm instead of 1.4 $\mu$m between the substrate and the QW. If the local density inhomogeneities originate from the residual charge patches at the substrate-epitaxy interface, the decreased buffer-layer thickness should amplify the effect and cause larger 2DES density fluctuations. Figure 3(a) shows the density map of the sample with a buffer-layer thickness of 200 nm. Compared to the sample with 1.4 $\mu$m buffer-layer thickness shown in Fig. 2(d), it is clear that the intensity of local density inhomogeneities is much more pronounced. We can therefore rule out oval defects as the primary source for the observed fluctuations as the oval defect density is known to increase with increasing thickness of the epitaxially-grown layer \cite{Oval}.

For quantitative comparison, in Fig. 3(c) we show line profiles of the contour plots for the two samples with different buffer-layer thicknesses. The black (red) dotted lines in Fig. 3(c) show data along the $x$-axis for $y=100$ (900) $\mu$m in Fig. 2(d), while the colored solid lines show profiles along the corresponding dashed lines shown in Fig. 3(a). Given that the spacer thickness and QW width of the two samples are identical, we expect the 2DES densities to also be quite similar in the two cases. Indeed, the median density values of the two samples are both $\sim2.95\times10^{11}$ cm$^{-2}$. Keeping this in mind, the data in Fig. 3(c) then reveal that the intensity of density fluctuations increases significantly more for depletion than accumulation when the buffer-layer thickness is decreased. In fact, in the sample with the thicker buffer layer [Fig. 2(d)], locally depleted regions show a density drop of $\sim1.2\times10^9$ cm$^{-2}$ while the depleted regions in the thin-buffer sample [Fig. 3(a)] show a density drop of $\sim9\times10^{9}$ cm$^{-2}$, which is in excellent agreement with the factor of 7 change in the buffer-layer thickness. We note that the low-temperature ($T=0.3$ K) transport mobility shows a significant drop of $\sim32\%$ to $\mu=4.4\times10^6$ cm$^{2}$V$^{-1}$s$^{-1}$ from $\mu=6.3\times10^6$ cm$^{2}$V$^{-1}$s$^{-1}$ when the buffer-layer thickness is decreased. This provides evidence that there is a clear correlation between the local density fluctuations and the 2DES mobility.

The linear scaling of the amount of charge depleted as the buffer-layer thickness is decreased strongly supports our conjecture that charged defects are located at the substrate-epitaxy interface and influence the local density structure of a 2DES in a GaAs QW. For the case of depletion, we suspect that C atoms are accountable for this type of density inhomogeneity as they are known to act as acceptors in GaAs \cite{Carbonacceptor1,Carbonacceptor2}. Given that the separation between the depleted regions is on the order of $\sim100$ $\mu$m, it is worthwhile to also consider dislocations in the GaAs substrate as a possible source of the local depletion. The specified etch-pit density of wafers typically used for growth in our chamber is $1500\sim5000$ cm$^{-2}$, which translates to a length scale of $140\sim250$ $\mu$m. This is comparable to the separation length mentioned above, and we cannot disregard the possibility that acceptor-like charges form at the dislocation sites near the substrate-epitaxy interface and contribute to the depletion observed in our data. 

Despite the prominent changes in the degree of depletion as the buffer-layer thickness was decreased, there seems to be much less of an effect for accumulation. As shown in Fig. 3(c), in general the amount of accumulated charge does not show a significant difference for the two samples. Compared to the median value of $\sim2.95\times10^{11}$ cm$^{-2}$, in accumulated regions the thin- and thick-buffer-layer samples typically show an increase of $\sim2.7\times10^9$ cm$^{-2}$ and $\sim2.4\times10^9$ cm$^{-2}$, respectively. The density change ratio of 1.13 does not coincide with the factor of 7 decrease in the buffer-layer thickness, which suggests that the accumulation is not related to trapped charges at the substrate-epitaxy interface. We see similar weak and broad accumulation patches even in samples with a buffer-layer thickness of 7 $\mu$m, further strengthening this argument (data not shown). At the moment we are unsure of what causes this type of density fluctuations. As mentioned earlier, it is possible that it is caused by impurities at the GaAs wafer surface or background impurities in the epitaxially grown structure.

Following the discussion in the previous paragraphs, it is interesting to investigate the change in the 2DES density map after the sample experiences a thermal cycle. If the local charge fluctuations in Fig. 3(a) come from fixed charge sources trapped at the substrate-epitaxy interface or other residual impurities in the sample, warming the system up to room temperature and then cooling it back down should have minimal impact on the overall features in the density map. Figure 3(b) shows the density map of the same region probed in Fig. 3(a) after the sample was warmed up to 300 K and cooled down to the base temperature again without breaking vacuum in the sample space. The data appear quite reproducible, showing similar density fluctuations in terms of both position and intensity even after a full thermal cycle. 


In conclusion, we have demonstrated a technique that can map out the local density of a 2DES using scanning PL. A GaAs sample with an intended density gradient of $\sim1$\%/mm was grown by MBE and was verified by our PL setup to have a density drop of $\sim4.5$\% over a 4 mm piece. In addition to this expected global variation in 2DES density, the density map revealed other, more local, density fluctuations which were also found in a sample grown without any intentional density gradients. We attribute the locally depleted regions to residual acceptor-like impurities at the substrate-epitaxy interface, and our conjecture is supported by the fact that the average local depletion increased from $\sim1.2\times10^9$ cm$^{-2}$ to $\sim9\times10^{9}$ cm$^{-2}$ when the buffer-layer thickness was decreased from 1.4 $\mu$m to 200 nm. The position and magnitude of the  density fluctuations we observe are quite reproducible even after a full thermal cycle to room temperature, which is consistent with the fact that residual impurities trapped at the substrate-epitaxy interface should not be very mobile. Our data directly show that there are local density inhomogeneities on the order of $\sim10^9$ cm$^{-2}$ even in clean 2DESs, quantitatively corroborating previous speculations based on anomalous features in transport measurements \cite{LMR1,quantumlifetime}. The technique and results presented here should find use in various topics of 2DES research that cannot be probed by transport measurements on macroscopic samples.


\begin{acknowledgments}
We acknowledge support through the NSF (Grants DMR 1709076 and ECCS 1508925) and the Department of Energy Basic Energy Sciences (DEFG02-00-ER45841) for measurements, and the NSF (Grant MRSEC DMR 1420541), the Gordon and Betty Moore Foundation (Grant GBMF4420) for sample fabrication and characterization.
 \end{acknowledgments}

 


\begin{thebibliography}{99}

\bibitem{Stormer} Dingle, R.; St\"{o}rmer, H. L.; Gossard, A. C.; Wiegmann, W. Electron mobilities in modulation-doped semiconductor heterojunction superlattices. \textit{Appl. Phys. Lett.} {\bf 1978}, \textit{33}, 665-667.

\bibitem{Gossard} English, J. H.; Gossard, A. C.; St\"{o}rmer, H. L.; Baldwin, K. W. GaAs structures with electron mobility of $5\times10^{6}$ cm$^2$/Vs. \textit{Appl. Phys. Lett.} {\bf 1987}, \textit{50}, 1826-1828.

\bibitem{Shayegan.APL} Shayegan, M.; Goldman, V. J.; Jiang, C.; Sajoto, T.; Santos, M. Growth of low-density two-dimensional electron system with very high mobility by molecular beam epitaxy. \textit{Appl. Phys. Lett.} {\bf 1988}, \textit{52}, 1086-1088.

\bibitem{Pfeiffer.APL} Pfeiffer, L.;  West, K. W.; Stormer, H. L.; Baldwin, K. W. Electron mobilities exceeding $10^7$ cm$^2$/Vs in modulation-doped GaAs. \textit{Appl. Phys. Lett.} {\bf 1989}, \textit{55}, 1888-1890.

\bibitem{DW1} Pfeiffer, L.; West, K. W. The role of MBE in recent quantum Hall effect physics discoveries. \textit{Physica E} {\bf 2003}, \textit{20}, 57-64.


\bibitem{Tsui.PRL.1982} Tsui, D. C.; Stormer, H. L.; Gossard, A. C. Two-dimensional magnetotransport in the extreme quantum limit. \textit{Phys. Rev. Lett.} {\bf 1982}, \textit{48}, 1559-1562.

\bibitem{QPC} van Wees, B. J.; van Houten, H.; Beenakker, C. W. J.; Williamson, J. G.; Kouwenhoven, L. P.; van der Marel, D.; Foxon, C. T. Quantized conductance of point contacts in a two-dimensional electron gas. \textit{Phys. Rev. Lett.} {\bf 1988}, \textit{60}, 848-850.

\bibitem{QPC2} Beenakker, C.W.J.; van Houten, H.; Quantum transport in semiconductor nanostructures. \textit{Solid State Phys.} {\bf 1991}, \textit{44}, 1-228.


\bibitem{Wigner1} Andrei, E. Y.; Deville, G.; Glattli, D. C.; Williams, F. I. B.; Paris, E.; Etienne, B. Observation of a magnetically induced Wigner solid. \textit{Phys. Rev. Lett.} {\bf 1988}, \textit{60}, 2765-2768.

\bibitem{Wigner2} Jiang, H. W.; Willett, R. L.; Stormer, H. L.; Tsui, D. C.; Pfeiffer, L. N.; West, K. W. Quantum liquid versus electron solid around $\nu=1/5$ Landau-level filling. \textit{Phys. Rev. Lett.} {\bf 1990}, \textit{65}, 633-636.

\bibitem{Wigner3} Goldman, V. J.; Santos, M.; Shayegan, M.; Cunningham, J. E. Evidence for two-dimentional quantum Wigner crystal. \textit{Phys. Rev. Lett.} {\bf 1990}, \textit{65}, 2189-2192.

\bibitem{stripebubble} Shayegan, M. Flatland electrons in high magnetic fields. In {\it High Magnetic Fields: Science and Technology}, edition 1; Herlach, F.; Miura, N., Eds.; World Scientific: Singapore, 2006; Vol. 3, pp. 31-60.

\bibitem{Fradkin.Review} Fradkin, E.; Kivelson, S. A.; Lawler, M. J.; Eisenstein, J. P.; Mackenzie, A. P. Nematic Fermi fluids in condensed matter physics. \textit{Annu. Rev. Condens. Matter Phys.} {\bf 2010}, \textit{1}, 153-178.

\bibitem{Si1} Lu, T. M.; Pan, W.; Tsui, D. C.; Lee, C.-H.; Liu, C. W. Fractional quantum Hall effect of two-dimensional electrons in high-mobility Si/SiGe field-effect transistors. \textit{Phys. Rev. B} {\bf 2012}, \textit{85}, 121307(R).

\bibitem{Si2} Kott, T. M.; Hu, Binhui; Brown, S. H.; Kane, B. E. Valley-degenerate two-dimensional electrons in the lowest Landau level. \textit{Phys. Rev. B} {\bf 2014}, \textit{89}, 041107(R).

\bibitem{Etienne.APL} De Poortere, E. P.; Shkolnikov, Y. P.; Tutuc, E.; Papadakis, S. J.; Shayegan, M.; Palm, E.; Murphy, T. Enhanced electron mobility and high order fractional quantum Hall states in AlAs quantum wells. \textit{Appl. Phys. Lett.} {\bf 2002}, \textit{80}, 1583-1585.

\bibitem{Chung.PRM} Chung, Y. J.; Villegas Rosales, K. A.; Deng, H.; Baldwin, K. W.; West, K. W.; Shayegan, M.; Pfeiffer, L. N. Multivalley two-dimensional electron system in an AlAs quantum well with mobility exceeding $2\times10^6$ cm$^{2}$V$^{-1}$s$^{-1}$. \textit{Phys. Rev. Materials} {\bf 2018}, \textit{2}, 071001R.

\bibitem{GaN} Manfra, M. J.; Weimann, N. G.; Hsu, J. W. P.; Pfeiffer, L. N.; West, K. W.; Syed, S.; Stormer, H. L.; Pan, W.; Lang, D. V.; Chu, S. N. G.; Kowach, G.; Sergent, A. M.; Caissie, J.; Molvar, K. M.; Mahoney, L. J.; Molnar, R. J. High mobility AlGaN/GaN heterostructures grown by plasma-assisted molecular beam epitaxy on semi-insulating GaN templates prepared by hydride vapor phase epitaxy. \textit{J. Appl. Phys.} {\bf 2002}, \textit{92}, 338-345.

\bibitem{CdTe} Piot, B. A.; Kunc, J.; Potemski, M.; Maude, D. K.; Betthausen, C.; Vogl, A.; Weiss, D.; Karczewski, G.; Wojtowicz, T. Fractional quantum Hall effect in CdTe. \textit{Phys. Rev. B} {\bf 2010}, \textit{82}, 081307(R).

\bibitem{ZnO} Falson, J.; Tabrea, D.; Zhang, D.; Sodemann, I.; Kozuka, Y.; Tsukazaki, A.; Kawasaki, M.; von Klitzing, K.; Smet, J. H. A cascade of phase transitions in an orbitally mixed half-filled Landau level. \textit{Sci. Adv.} {\bf 2018}, \textit{4}, eaat8742.

\bibitem{Ge} Shi, Q.; Zudov, M. A.; Morrison, C.; Myronov, M. Spinless composite fermions in an ultrahigh-quality strained Ge quantum well. \textit{Phys. Rev. B} {\bf 2015}, \textit{91}, 241303(R).

\bibitem{InAs} Ma, Meng K.; Hossain, Md. Shafayat; Villegas Rosales, K. A.; Deng, H.; Tschirky, T.; Wegscheider, W.; Shayegan, M. Observation of fractional quantum Hall effect in an InAs quantum well. \textit{Phys. Rev. B} {\bf 2017}, \textit{96}, 241301(R).

\bibitem{Zibrov.NatPhys} Zibrov, A. A.; Spanton, E. M.; Zhou, H.; Kometter, C.; Taniguchi, T.; Watanabe, K.; Young, A. F.; Even-denominator fractional quantum Hall states at an isospin transition in monolayer graphene. \textit{Nat. Phys.} {\bf 2018}, \textit{14}, 930-935.

\bibitem{Cao.Nature1} Cao, Y.; Fatemi, V.; Demir, A.; Fang, S.; Tomarken, S. L.; Luo, J. Y.; Sanchez-Yamagishi, J. D.; Watanabe, K.; Taniguchi, T.; Kaxiras, E.; Ashoori, R. C.; Jarillo-Herrero, P. Correlated insulator behaviour at half-filling in magic-angle graphene superlattices. \textit{Nature} {\bf 2018}, \textit{556}, 80-84.

\bibitem{Cao.Nature2} Cao, Y.; Fatemi, V.; Fang, S.; Watanabe, K.; Taniguchi, T.; Kaxiras, E.; Jarillo-Herrero, P. Unconventional superconductivity in magic-angle graphene superlattices. \textit{Nature} {\bf 2018}, \textit{556}, 43-50.





\bibitem{ManfraGa} Gardner, G. C.; Fallahi, S.; Watson, J. D.; Manfra, M. J. Modified MBE hardware and techniques and role of gallium purity for attainment of two dimensional electron gas mobility $>35\times10^6$ cm$^2$/Vs in AlGaAs/GaAs quantum wells grown by MBE. \textit{J. Cryst. Growth} {\bf 2016}, \textit{441}, 71-77.

\bibitem{Gaproblem} Schlapfer, F.; Dietsche, W.; Reichl, C.; Faelt, S.; Wegscheider, W. Photoluminescence and the gallium problem for highest-mobility GaAs/AlGaAs-based 2d electron gases. \textit{J. Cryst. Growth} {\bf 442}, \textit{442}, 114-120.

\bibitem{Alproblem} Chung, Y. J.; Baldwin, K. W.; West, K. W.; Shayegan, M.; Pfeiffer, L. N. Surface segregation and the Al problem in GaAs quantum wells. \textit{Phys. Rev. Materials} {\bf 2018}, \textit{2}, 034006 (2018).

\bibitem{LMR1} Pan, W.; Xia, J. S.; Stormer, H. L.; Tsui, D. C.; Vicente, C. L.; Adams, E. D.; Sullivan, N. S.; Pfeiffer, L. N.; Baldwin, K. W.; West, K. W. Quantization of the diagonal resistance: density gradients and the empirical resistance rule in a 2D System. \textit{Phys. Rev. Lett.} {\bf 2005}, \textit{95}, 066808.

\bibitem{DensityInhomogeneity} Samani, M.; Rossokhaty, A. V.; Sajadi, E.; L\"{u}scher, S.; Folk, J. A.; Watson, J. D.; Gardner, G. C.; Manfra, M. J. Low-temperature illumination and annealing of ultrahigh quality quantum wells. \textit{Phys. Rev. B} {\bf 2014}, \textit{90}, 121405(R).

\bibitem{Map} Zhou, W.; Yoo, H. M.; Prabhu-Gaunkar, S.; Tiemann, L.; Reichl, C.; Wegscheider, W.; Grayson, M. Analyzing longitudinal magnetoresistance asymmetry to quantify doping gradients: generalization of the van der Pauw method. \textit{Phys. Rev. Lett.} {\bf 2015}, \textit{115}, 186804.

\bibitem{LMR2} Khouri, T.; Zeitler, U.; Reichl, C.; Wegscheider, W.; Hussey, N. E.; Wiedmann, S.; Maan, J. C. Linear magnetoresistance in a quasifree two-dimensional electron gas in an ultrahigh mobility GaAs quantum well. \textit{Phys. Rev. Lett.} {\bf 2016}, \textit{117}, 256601.

\bibitem{quantumlifetime} Qian, Q.; Nakamura, J.; Fallahi, S.; Gardner, G. C.; Watson, J. D.; L\"{u}scher, S.; Folk, J. A.; Cs\'{a}thy, G. A.; Manfra, M. J. Quantum lifetime in ultrahigh quality GaAs quantum wells: relationship to $\Delta_{5/2}$ and impact of density fluctuations. \textit{Phys. Rev. B} {\bf 2017}, \textit{96}, 035309.

\bibitem{Kamburov.APL} Kamburov, D.; Baldwin, K. W.; West, K. W.; Lyon, S.; Pfeiffer, L. N.; Pinczuk, A. Use of micro-photoluminescence as a contactless measure of the 2D electron density in a GaAs quantum well. \textit{Appl. Phys. Lett.} {\bf 2017}, \textit{110}, 262104.

\bibitem{BGR} Kleinman, D. A.; Miller, R. C. Band-gap renormalization in semiconductor quantum wells containing carriers. \textit{Phys. Rev. B} {\bf 1985}, \textit{32}, 2266-2272.

\bibitem{Saito.J.Crys.Growth} Saito, J.; Nanbu, K.; Ishikawa, T.; Kondo, K. In situ cleaning of GaAs substrates with HCl gas and hydrogen mixture prior to MBE growth. \textit{J. Cryst. Growth} {\bf 1989}, \textit{95}, 322-327.

\bibitem{Lee.J.Crys.Growth} Lee, J.J.D.; West, K.W.; Baldwin, K.W.; Pfeiffer, L.N. Smoothness and cleanliness of the GaAs (100) surface after thermal desorption of the native oxide for the synthesis of high mobility structures using molecular beam epitaxy. \textit{J. Cryst. Growth} {\bf 2012}, \textit{356}, 46-52.

\bibitem{Oval} Wang, Y. H.; Liu, W. C.; Chang, C. Y.; Liao, S. A. Surface morphologies of GaAs layers grown by arsenic-pressure-controlled molecular
beam epitaxy. \textit{J. Vac. Sci. Technol. B} {\bf 1986}, \textit{4}, 30-36.


\bibitem{Carbonacceptor1} de Lyon, T. J.; Woodall, J. M.; Goorsky, M. S.; Kirchner, P. D. Lattice contraction due to carbon doping of GaAs grown by metalorganic molecular beam epitaxy. \textit{Appl. Phys. Lett.} {\bf 1990}, \textit{56}, 1040-1042.

\bibitem{Carbonacceptor2} Stockman, S. A.; H\"{o}fler, G. E.; Baillargeon, J. N.; Hsieh, K. C.; Cheng, K. Y.; Stillman, G. E. Characterization of heavily carbon-doped GaAs grown by metalorganic chemical vapor deposition and metalorganic molecular beam epitaxy. \textit{J. Appl. Phys.} {\bf 1992}, \textit{72}, 981-987.


\end{thebibliography}
\end{document}